\documentclass[twoside,a4paper,11pt]{sca}
\usepackage{graphicx}
\usepackage{hyperref}
\usepackage{movie15}
\usepackage{natbib}  
\begin{document}
\pagenumbering{arabic}
\pagestyle{myheadings}
\thispagestyle{empty}
{\flushright\includegraphics[width=\textwidth,bb=90 650 520 700]{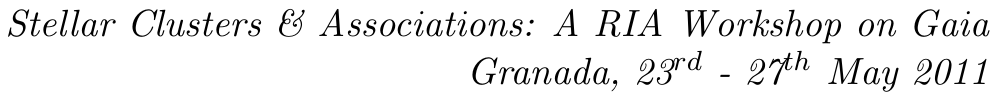}}
\vspace*{0.2cm}
\begin{flushleft}
{\bf {\LARGE
%
Using open clusters to study mixing in low- and intermediate-mass stars
%
}\\
\vspace*{1cm}
%
Rodolfo Smiljanic
%
}\\
\vspace*{0.5cm}
%
European Southern Observatory, Garching bei M\"unchen, Germany \\
%
\end{flushleft}
%
\markboth{
Mixing in stars of open clusters
}{ 
%
R. Smiljanic
%
}
\thispagestyle{empty}
\vspace*{0.4cm}
\begin{minipage}[l]{0.09\textwidth}
\ 
\end{minipage}
\begin{minipage}[r]{0.9\textwidth}
\vspace{1cm}
\section*{Abstract}{\small
%
In many evolutionary stages, low- and intermediate-mass stars show signs of mixing of the surface material with 
material from the interior. To account for all the details revealed by the observations it is necessary to include 
non-standard physical processes in the models (e.g. atomic diffusion and rotation-induced mixing). The 
study of mixing in stars of different masses, ages, and chemical composition helps to identify 
and constrain these processes. In this sense, stars in open clusters are the ideal targets. All stars in one given cluster have the same age 
and chemical composition, and their masses can be well estimated. By studying many clusters, one can separate and trace the effects of these 
different parameters.



%
\normalsize}
\end{minipage}
%
%
%
\section{The early main sequence}

Lithium and beryllium burn at low but different temperatures (2.5 $\times$ 10$^6$ K and 3.5 $\times$ 10$^6$ K, respectively). Their 
abundances can be used to study mixing in the outer layers of stars \citep[see e.g.][and references therein]{Sm10}. Observations 
of stars in young clusters (from $\sim$ 10 to 100 Myr) have shown that pre-main sequence (PMS) models predict too 
much Li depletion. In addition, there is a significant spread in Li abundances among 
late-G and K dwarfs which is connected to rotation, i.e. fast-rotating stars have higher Li abundances than slow-rotating ones 
\citep[see e.g.][and references therein]{Balachandran11}. 

Beryllium has been studied in far fewer stars than Li has. In Smiljanic, Randich \& Pasquini (2011, submitted) we derived Be abundances 
in ten G- and K-type main-sequence stars of the open clusters IC 2391 and IC 2602 ($\sim$ 50 Myr). As these stars have just arrived on 
the main sequence, any change in Li and Be should have taken place during the PMS.

All stars have, within the uncertainties, the same Be abundances even though their Li abundances differ by almost 
one order of magnitude. This confirms what is expected by the models; Be abundances are not affected by mixing during the 
PMS in stars between 0.80 $\leq$ M/M$_{\odot}$ $\leq$ 1.20. Comparing 
our Be abundances with those of other young clusters from the literature, we confirm that Be depletion in stars 
cooler than $\sim$ 5600 K increases with age. The depletion is stronger for smaller masses. As stars with ages of 50 and 150 Myr do not 
show Be depletion, we conclude that the depletion in older stars happens during the main sequence, in agreement with \citet{Ran07}.

\section{The Li-Be dip}

The so-called Li-dip is a strong decrease in the Li abundance seen in main-sequence stars in a small temperature range around 6700 K. 
It was first found by \citet{BT86a} in the Hyades. Be was also found to be depleted in these stars \citep[see e.g.][and references therein]{Sm10}.

In \citet{Sm10} we derived Be abundances along the whole evolutionary sequence of the cluster IC 4651 ($\sim$ 1.7 Gyr). A well defined LiBe-dip was 
found \citep[Li abundances from][]{PRZ04}. The hydrodynamical models of \citet{CharbonnelLagarde10} reproduce well the observed 
behavior of Li and Be over the whole temperature range. The hot side of the dip requires models with atomic diffusion, and transport of angular momentum and 
chemicals by meridional circulation and shear turbulence. In the cool side of the dip, the models also take into account extraction of angular momentum by internal gravity waves.

\section{The red giant branch}

Abundances of C, N, O, and the $^{12}$C/$^{13}$C in low-mass giants, before the bump were found to be in good agreement with theoretical 
predictions. However, after the bump an additional mixing event takes place resulting in further abundance changes \citep[see e.g.][and 
references therein]{Sm09b}.

In \citet{Sm09b} we derived C, N, O, Na, and $^{12}$C/$^{13}$C in 31 giants, with 1.7 $\leq$ M/M${\odot}$ $\leq$ 3.1, of 10 open clusters. We 
found that the well-known trend of decreasing carbon ratio with decreasing mass is not so well defined, but shows a significant 
scatter.  The decrease of $^{12}$C/$^{13}$C can be explained by the action of thermohaline mixing while the scatter might be 
partially explained by a dispersion in the initial rotation velocities of the stars \citep{CharbonnelLagarde10}. The lowest values found 
are however difficult to explain. New homogeneous analyses are needed to confirm and better 
constrain these results.

%
%
\small  
%
%

%
%
%
%
%

\bibliographystyle{aa}
\bibliography{ref_smiljanic_r}

\begin{thebibliography}{7}
\expandafter\ifx\csname natexlab\endcsname\relax\def\natexlab#1{#1}\fi

\bibitem[{{Balachandran} {et~al.}(2011){Balachandran}, {Mallik}, \&
  {Lambert}}]{Balachandran11}
{Balachandran}, S.~C., {Mallik}, S.~V., \& {Lambert}, D.~L. 2011, MNRAS, 410,
  2526

\bibitem[{{Boesgaard} \& {Tripicco}(1986)}]{BT86a}
{Boesgaard}, A.~M. \& {Tripicco}, M.~J. 1986, ApJL, 302, L49

\bibitem[{{Charbonnel} \& {Lagarde}(2010)}]{CharbonnelLagarde10}
{Charbonnel}, C. \& {Lagarde}, N. 2010, A\&A, 522, A10+

\bibitem[{{Pasquini} {et~al.}(2004){Pasquini}, {Randich}, {Zoccali}, {Hill},
  {Charbonnel}, \& {Nordstr{\"o}m}}]{PRZ04}
{Pasquini}, L., {Randich}, S., {Zoccali}, M., {et~al.} 2004, A\&A, 424, 951

\bibitem[{{Randich} {et~al.}(2007){Randich}, {Primas}, {Pasquini}, {Sestito},
  \& {Pallavicini}}]{Ran07}
{Randich}, S., {Primas}, F., {Pasquini}, L., {Sestito}, P., \& {Pallavicini},
  R. 2007, A\&A, 469, 163

\bibitem[{{Smiljanic} {et~al.}(2009){Smiljanic}, {Gauderon}, {North}, {Barbuy},
  {Charbonnel}, \& {Mowlavi}}]{Sm09b}
{Smiljanic}, R., {Gauderon}, R., {North}, P., {et~al.} 2009, A\&A, 502, 267

\bibitem[{{Smiljanic} {et~al.}(2010){Smiljanic}, {Pasquini}, {Charbonnel}, \&
  {Lagarde}}]{Sm10}
{Smiljanic}, R., {Pasquini}, L., {Charbonnel}, C., \& {Lagarde}, N. 2010, A\&A,
  510, A50

\end{thebibliography}

\end{document}